\documentclass[lncs,runningheads,a4paper]{llncs}

\usepackage{etex}
\usepackage{hyperref}
\usepackage[utf8]{inputenc}
\usepackage[T1]{fontenc}
\usepackage[ngerman,english]{babel}
\usepackage{times}
\usepackage{paralist}
\usepackage{graphicx}
\usepackage[dvipsnames]{xcolor}
\usepackage{caption}
\usepackage{subcaption}
\usepackage{booktabs}
\usepackage{amssymb, amsfonts}
\usepackage{amsmath}
\usepackage{amsopn}
\usepackage{wrapfig}
\usepackage{nicefrac}

\usepackage{tikz, tikz-qtree}
\usepackage[ruled,linesnumbered,vlined]{algorithm2e}
\usepackage{microtype}
\usepackage{url}
\usepackage{balance}

\usepackage{multirow}
\usepackage{relsize}
\usepackage{caption}
\usepackage{cite}

\graphicspath{{./fig/}}

\SetCommentSty{mycommfont}

\DeclareMathOperator{\dom}{dom}

\newcommand{\mytitle}{Privacy-aware Process Performance Indicators:\\ 
Framework and Release Mechanisms }

\newcommand{\myauthorhyperref}{Kabierski et al.}

\hypersetup{
	bookmarks=true,        % show bookmarks bar?
	unicode=true,          % non-latin characters in bookmarks?
	pdffitwindow=false,    % window fit to page when opened
	pdfstartview={FitH},   % fits width of the page to the window
	pdftoolbar=true,       % show acrobat toolbar?
	pdfmenubar=true,       % show acrobat menu?
	colorlinks=true,      % false: boxed links; true: color links
	linkcolor={blue!80!black},       % color of internal links
	urlcolor={blue!30!black},        % color of external links
	citecolor={green!60!black},       % color of links to bibliography
	filecolor=black,       % color of file links
	pdftitle={\mytitle},           % title
	pdfauthor={\myauthorhyperref}, % author
	pdfcreator={\myauthorhyperref} % creator of pdf
}

\begin{document}
\title{\mytitle}

\author{Martin Kabierski \and Stephan A. 
Fahrenkrog-Petersen\and  Matthias 
Weidlich}
\authorrunning{M. Kabierski et al.}
\titlerunning{Privacy-aware Process Performance Indicators}
\institute{Humboldt-Universität zu Berlin, Germany\\
\email{martin.bauer | stephan.fahrenkrog-petersen | 
matthias.weidlich@hu-berlin.de}
}
\maketitle              % typeset the header of the contribution
\begin{abstract}
Process performance indicators (PPIs) are metrics to quantify the degree with  
which organizational goals defined based on business processes are 
fulfilled. They exploit the event logs recorded by information systems during 
the execution of business processes, thereby providing a basis for process 
monitoring and subsequent optimization. 
However, PPIs are often evaluated on processes that involve individuals, which 
implies an inevitable risk of privacy intrusion. 
In this paper, we address the demand for privacy protection in the computation 
of PPIs. We first present a framework that enforces control over the 
data exploited for process monitoring. We then show how PPIs defined based on 
the established PPINOT meta-model are instantiated in this framework through 
a set of data release mechanisms. These mechanisms are designed to provide 
provable guarantees in terms of differential privacy. 
%Yet, up to this point, no 
%privacy protection model for the evaluation of PPIs has been proposed in 
%literature. To this end, we aim to close this gap, by introducing the first 
%protection model for the evaluation of PPIs defined using the PPINOT 
%metamodel. 
%We propose a centralized architecture and a set of provably 
%$\epsilon$-differentially private release mechanisms. 
We evaluate our framework and the release mechanisms in a series of 
controlled experiments. We further use a public event log to compare our framework with approaches based on privatization of event logs.
The results demonstrate feasibility and shed light on the trade-offs 
between data utility and privacy guarantees in the computation of PPIs. 

\end{abstract}

	\section{Introduction}
	\vspace{-.5em}
	\label{sec:introduction}
	% !TeX spellcheck = en_GB
% !TeX root = ../main.tex

\begin{figure}[t!]
	\centering
	\begin{subfigure}[b]{0.5\textwidth}
		\centering
		\includegraphics[width=\columnwidth]{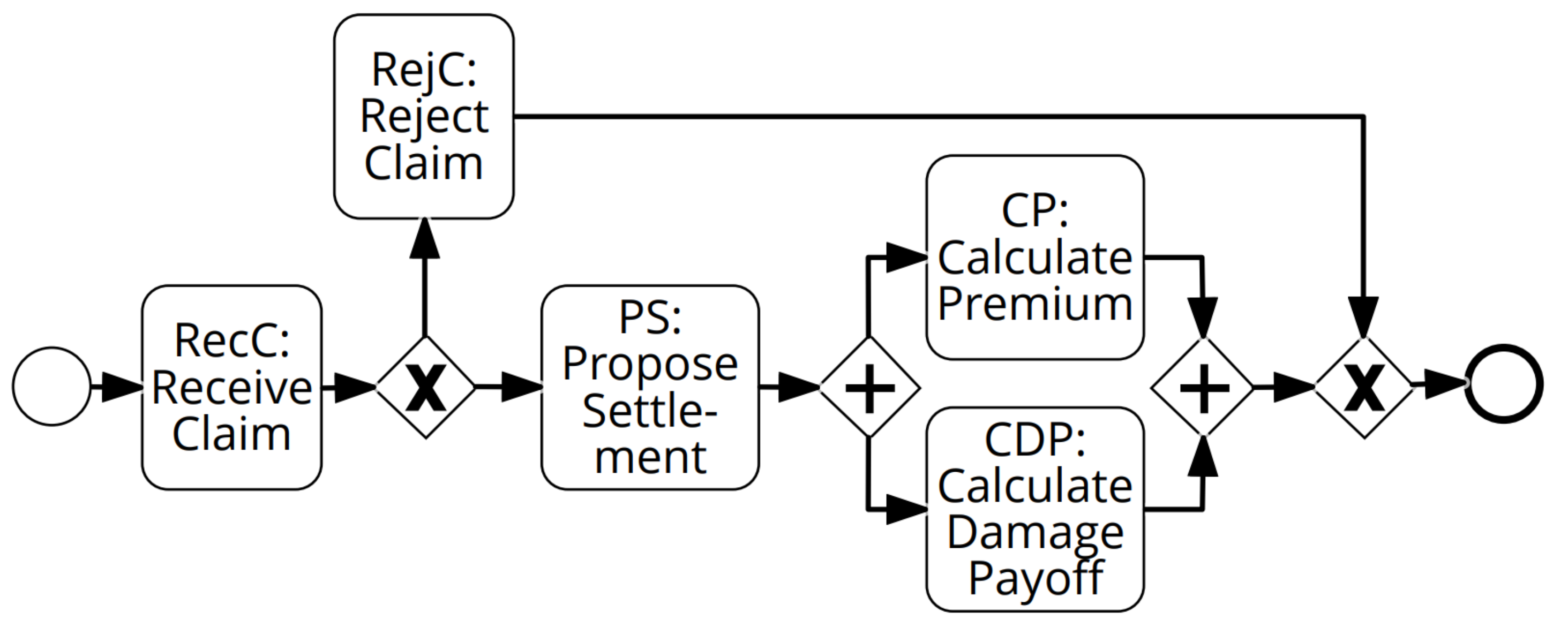}
		\caption{}
	\label{fig:model}
	\end{subfigure}
	\begin{subfigure}[b]{0.38\textwidth}
		\scriptsize
		\centering
		\begin{tabular}{l @{\quad } l @{\quad } l}
			\toprule
			Measure & Target & Scope\\
			\midrule
			1: Mean sojourn time & <1 week & Month\\
			2: \% rejected claims & >80\% & Month\\
			3: Mean damage pay. & <5.000$\$$ & Week\\
%			5:& Fraction of damage payment > 20.000$\$$ & <5\% & Monthly\\
			\bottomrule
		\end{tabular}
		\caption{}
		\label{fig:PPI}
	\end{subfigure}
	\vspace{-.2em}
	\caption{(a) Model of a claim handling process; (b) PPIs defined on the 
	model.}
	\label{fig:running}
	\vspace{-.25em}
\end{figure}

%\begin{compactitem}
%\item motivation
%\item small example process including performance indicators, this needs to be 
%very concise, model from Fig 1 in the thesis (just much more compact) and table 
%from Fig 3
%\item research question
%\item contributions
%\item evaluation highlights
%\end{compactitem}
%\todoinline{R1M, R3C4: Update repository}
Many companies improve their operation by applying process-oriented 
methodologies. In this context, Business Process Management (BPM) provides 
methods and techniques to aid in the
monitoring, analysis, and optimization of business 
processes~\cite{DBLP:books/sp/DumasRMR18}. Important means to enable
the continuous optimization of processes are \emph{process 
performance indicators} (PPIs), 
i.e. numerical measures computed based on data recorded during process 
execution~\cite{delrioortega2013}. PPIs assess whether predefined goals set by 
the process owner are fulfilled, e.g., related to the mean sojourn time of a 
business process. \autoref{fig:running} illustrates a simple insurance claim 
handling process and respective PPIs. Each indicator comprises a definition of 
a measure, a target value, and an observation period, called scope.
 % a scope describing the time frame, in which these PPI shall be evaluated.

The data used to calculate PPIs often includes personal data. In 
\autoref{fig:running}, such data relates to the knowledge workers handling 
the claims or the customers who submitted them. Processing of personal data is 
strictly regulated. The GDPR~\cite{eugdpr2018}, as an example, prohibits the use
of personal data without explicit consent and especially restricts their 
\emph{secondary use}, i.e., the processing of data beyond the purpose for which 
they were originally recorded. Process optimization typically represents such a 
secondary use of process execution data~\cite{MannhardtPO18}. To motivate, why 
unregulated access to process execution data may be problematic, we turn back 
to the example model and PPI in \autoref{fig:PPI}. Assume that data is recorded 
about three claims handled by Alice with sojourn times of 4, 4, and 5 days; 
three claims handled by Bob within 2, 6, and 6 days; and three claims handled 
by Sue lasting for 7, 8, and 8 days. Here, the mean sojourn time of these nine 
process instances is \textasciitilde 5.5 days and thus fulfils PPI 1 set by 
management. Yet, considering this data directly would reveal Sue's generally 
slower processing times, which may be prohibited by privacy regulations.
Here, privacy-protected PPI schemes, i.e., techniques that incorporate data 
anonymization in the computation of PPIs, would allow for the evaluation 
of PPIs, while protecting the privacy of the recorded individuals in the log 
file, thus lifting these privacy regulations. Yet, data anonymization commonly 
leads to a trade-off between the strength of a privacy guarantee and a loss in 
data utility, thus a privacy-protected PPI scheme needs to minimize the accuracy loss introduced.
%While data anonymization may lift such regulations, it commonly leads to a trade-off between the strength of a privacy guarantee and a loss in data utility. This calls for anonymization schemes that minimize the accuracy loss of PPI queries, so that management may still analyze the fulfilment of operational goals, while the privacy of involved individuals is protected.

Models for privacy-aware computation of traditional 
aggregates~\cite{aldeen2015, mendes2017} have limited applicability for PPIs, 
though. Since these models do not take into account the highly structured
nature of data generated by processes and PPIs defined on them, these methods are not suitable for privatizing PPIs. Approaches for privacy-aware publishing and querying 
of process execution data~\cite{fahrenkrog-petersen2019,mannhardt2019}, in
turn, are too coarse-grained. Handling comprehensive execution data, these 
techniques cannot be tailored to minimize the loss in data utility for a given 
set of PPIs. Against this background, we identify the research question of 
\textit{how to design a framework for the evaluation of 
privacy-protected PPIs.}

In this paper, we address the above question, by proposing  PaPPI, a first framework for 
\textbf{p}rivacy-\textbf{a}ware evaluation of \textbf{PPI}s. It separates trusted and untrusted 
environments to handle process execution data. They are connected by a 
dedicated interface that serves as a privacy checkpoint, ensuring $\epsilon$-differential privacy\cite{dwork2006}. We then instantiate 
this framework with data release mechanisms for PPIs that are defined based on 
the established PPINOT meta-model~\cite{delrioortega2013}. This way, we enable 
organizations to compute expressive PPIs without risking privacy violations. 
%Our model satisfies GDPR and allows organizations to utilize PPIs without 
%risking privacy violations.
%Furthermore, we will answer the following research questions:
%\begin{compactitem}
%	\item \textbf{RQ1}: How should a conceptual model for the privacy-protected 
%PPI evaluation be designed to allow for application in real world context?
%	\item \textbf{RQ2}: How can this model be instantiated with anonymization 
%mechanisms that allow for the unambiguous and expressive definition of PPIs 
%whilst providing provable privacy guarantees?
%	\item \textbf{RQ3}: Does the envisioned protection model under application 
%of the derived evaluation techniques allow for the evaluation of PPIs with 
%acceptable utility in real application contexts?
%\end{compactitem}
Finally, we explore the impact of privacy-aware evaluation of PPIs 
on their quality. We report on controlled experiments using 
synthetic data and a case study with a publicly available 
event log. Our results demonstrate the feasibility 
of the framework and its instantiation through specific release mechanisms, 
given that a reasonable amount of process execution data has been recorded.

In the remainder, \autoref{sec:background} provides background on PPIs and privacy guarantees. \autoref{sec:privacymodel} introduces our framework for privacy-aware 
evaluation of PPIs, which is instantiated 
with specific release mechanisms in \autoref{sec:evaluationtechniques} and 
evaluated in \autoref{sec:evaluation}. 
Finally, we review related work in \autoref{sec:relatedwork}, before we 
conclude in \autoref{sec:conclusion}.
	
	\section{Background}\vspace{-.5em}
	\label{sec:background}
	% !TeX spellcheck = en_GB
% !TeX root = ../main.tex

We introduce a basic model for event logs (\autoref{sec:notation}) and process 
performance indicators (\autoref{sec:ppi}). Finally, 
we review the concept of differential privacy %and related release mechanisms 
(\autoref{sec:differential_privacy}).

\subsection{Notions and Notations for Event Logs}
\label{sec:notation}
%\subsubsection{Basic notations}

We consider ordered, finite datasets, each being a set of 
elements $X=\{x_1, \ldots, x_n\}$ that carry a numeric value and are partially 
ordered by $\leq$.\footnote{For ease of presentation, we exemplify datasets as 
sets of 
integers or real numbers, even though in practice, a dataset may contain 
multiple elements referring to the same numeric value.} 
The cardinality of the dataset is denoted as $|X|=n$. For one of the 
(potentially many) elements of $X$ that are minimal and maximal according to 
$\leq$, we write 
$\underline{X}$ and $\overline{X}$, 
respectively. %For $X$, we get $\underline{X}=x_1$ and $\overline{X}=x_n$. 
An interval of the dataset is defined by $I=(x_{lower},x_{upper})$ with 
$x_{lower},x_{upper}\in X$ and $x_{lower}\leq x_{upper}$. 
Lifting the notation for minima and maxima to $I$, we define 
$\underline{I}=x_{lower}$ and $\overline{I}=x_{upper}$.

%We consider ordered, finite datasets, each being represented by a multi-set of 
%real values, $X=[x^{k_1}_1, \ldots, x^{k_n}_n]$ with ${k_i}$ as the 
%cardinality 
%of value $x_i$.
%%, ordered from lowest to the highest value, i.e. $X=\{x_1, x_2, ..., x_n\}$ 
%%with $x_1\leq x_2\leq ...\leq x_n$. 
%The cardinality of the dataset is denoted as $|X|=\sum_{i} k_i$. For the 
%minimum and 
%maximum value of $X$, we write $\underline{X}$ and $\overline{X}$, 
%respectively. %For $X$, we get $\underline{X}=x_1$ and $\overline{X}=x_n$. 
%With $I=(x_{lower},x_{upper})$ with $x_{lower},x_{upper}$ being values in $X$ 
%such that $x_{lower}\leq x_{upper}$, we denote an interval of the dataset. 
%Lifting the notation for minima and maxima to $I$, we define 
%$\underline{I}=x_{lower}$ and $\overline{I}=x_{upper}$.

%\subsubsection{Event Logs}
Our notion of an event log is based on a relational event 
model~\cite{arasu2016}. That is, an \emph{event schema} is defined by a tuple 
of attributes $A=(A_1,\ldots,A_n)$, so that an \emph{event} is an instance of 
the schema, i.e., a tuple of attribute values $e=(a_1,\ldots,a_n)$. An event 
schema 
consists of at least three attributes, the \textit{case} 
that identifies the process instance to which an event belongs, the 
\textit{timestamp} for the point in time an event has been recorded, and the 
\textit{activity}, for which the execution is signalled by an event. The 
\textit{timestamp}-ordered list of 
events corresponding to a single \textit{case} is called a \emph{trace}. Such a 
trace represents the execution of a single process instance. An \emph{event 
log} is a set of traces.

\subsection{Process Performance Indicators}
\label{sec:ppi}

A \emph{key performance indicator (KPI)} is a metric that quantifies, to which 
extent the goals set for an organisation are fulfilled. A \emph{process 
performance indicator (PPI)} is a KPI, which is related to a single business 
process and which is evaluated solely based on the traces recorded during 
process execution. The \emph{Process Performance Indicator Notation 
(PPINOT)}~\cite{delrioortega2013} is a meta-model for the 
definition and evaluation of PPIs. At its core, the PPINOT model 
relies on the composition of measures, i.e., simple, well-defined 
functions that enable the definition and automated evaluation of more complex 
PPIs:

 %For the calculation definition of the PPI, PPINOT uses pre-defined measure definitions, that resemble basic well-defined functions, that can be composed to calculate more complex functions. In particular, PPINOT considers these measure definitions to be separated into single-instance measures and multi-instance measures, that are evaluated on either a single process instance or a set of process instances. respectively. 
%These measure definitions are:
\begin{compactitem}
	\item[\textit{Base  measures}] concern a single instance of a 
	process and include event counts (e.g., to count activity executions), 
	timestamp differences between events, the satisfaction of conditions, or  
	aggregations over the events' attribute values. 
%    Therefore, these measures can be used for e.g. retrieving the 
%	time between two instants, counting how often a certain activity is executed 
%	in a process instance, evaluating whether a boolean condition holds, or 
%	quantifying the value of certain attributes in the process instance.
	\item[\textit{Aggregation measures}] are multi-instance measures that combine 
	values from multiple process instances into a single value. 
	PPINOT includes aggregation measures to calculate the minimum, maximum, mean, 
	and sum of a set of input values.
	\item[\textit{Derived measures}] are user-defined functions of arbitrary 
	form, applied to a single process instance, or a set thereof. 
\end{compactitem}

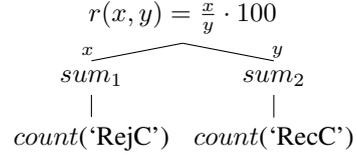
\begin{wrapfigure}{h}{0.4\columnwidth}
	\vspace{-.5em}
	\centering
	\begin{tikzpicture}[level distance=.8cm,
		level 1/.style={sibling distance=2.4cm},
		level 2/.style={sibling distance=1cm}]
		\node {$r(x,y) = \frac{x}{y}\cdot 100$}
		child {node {$sum_1$}
			child {node {$count$(`RejC')}}
		}
		child {node {$sum_2$}
			child {node {$count$(`RecC')}}
		};
		\node at (-1.25,-0.45) {\scriptsize $x$};
		\node at (1.25,-0.45) {\scriptsize $y$};
	\end{tikzpicture}
	\vspace{-.5em}
	\caption{The function composition tree of PPI 2 in \autoref{fig:PPI}.}
	\label{fig:function_composition_tree}
	\vspace{-2.5em}
\end{wrapfigure}

A PPI defined using the PPI meta-model is represented as a function 
composition tree. \autoref{fig:function_composition_tree} exemplifies such a 
tree for the PPI 2 of our example from \autoref{fig:PPI}. It calculates the 
fraction of rejected insurance claims and received 
claims. Here, \emph{count} is a base measure; \emph{sum} is an aggregation measure; and the fraction $r(x,y)$ is a derived measure. 
%First, two count measures are used to retrieve for 
%each instance, the number rejected Claims and Received Claims activities. 
%These are then aggregated using a Sum-measure.  As each process instance 
%starts with the receiving of a claim, this is equivalent to the sum of all 
%process instances in the scope. Finally, the number of rejected claims is 
%normalized by the number of total claims and multiplied by $100$, to get a 
%percentual value, in the final derived multi-instance measure. The other 
%four 
%PPIs can be represented in PPINOT as well in a similar manner.

%\todo{Recommendation from R1: Add a section before section 2.3, to discuss what we want to achieve with privacy-aware evaluation of PPIs}

\subsection{Differential Privacy}
\label{sec:differential_privacy}
Differential privacy~\cite{dwork2006} is a privacy guarantee  that limits the 
impact a single element may have on the output of a function $f$ that is 
computed over a set of elements. Therefore, it limits an adversary to conclude 
on the set of used input elements (or the presence of a certain input element) 
from the result of the function. This
obfuscation is usually achieved by adding noise, for which the magnitude 
depends on the sensitivity $\Delta f$ of function $f$, i.e., the 
%$\Delta f$ resembles the 
maximal impact any element $x\in X$ may have on $f(X)$.

A randomized mechanism $K$ is a randomized function that can be applied to a 
dataset, with $range(K)$ as the set of possible results. Let $D_1, D_2$ be two 
neighbouring datasets, i.e., they differ in exactly one element. 
The randomized mechanism $K$ provides $(\epsilon, \delta)$-differential 
privacy, if the following inequality holds for the probabilities of the 
function result falling into a sub-range of all possible results:
\begin{align*}
\forall \ S \subseteq range(K): \ P(K(D_1)\in  S)\ \leq\ e^\epsilon P(K(D_2)\in 
S)+ 
\delta
\end{align*}
Differential privacy enforces an upper bound on the difference in 
result probabilities of neighbouring datasets. If $\delta=0$, $K$ is 
$\epsilon$-differentially private (or $\epsilon$ is omitted altogether). 
Larger $\epsilon$ values imply weaker privacy, 
while the contrary holds true for 
smaller values. 

%A specific mechanisms to achieve differential privacy are the 
%Laplace mechanism~\cite{dwork2006}, the exponential 
%mechanism~\cite{mcsherry2007} and the sample-and-aggregate 
%framework~\cite{nissim2007}.

A specific mechanism to achieve differential privacy is the 
Laplace mechanism~\cite{dwork2006}. It adds noise 
sampled from a Laplace distribution with parameters $\mu=0$ and $b={\Delta 
f}/{\epsilon}$ onto $f(X)$. Due to the symmetric nature of the Laplacian and 
the exponential falloff, results are expected to lie close to 
$f(X)$. 

The symmetrical monotonous falloff of the Laplace 
mechanism may yield undesirable results, e.g., if values close to the true 
result have a disproportionally negative effect on the utility. The 
\emph{exponential mechanism}~\cite{mcsherry2007} avoids this problem, 
by constructing a probability space based on a function $q(D,r)$, 
which assigns a score to all results $r\in range(K)$ based on the input dataset 
$D$. Here, a higher score is assigned to more 
desirable results. %, which implies an exponentially higher result probability. 
The mechanism then chooses a result $r\in range(K)$ with a probability 
proportional to $e^{({\epsilon q(D,r)})/({2\Delta q})}$, where 
$\Delta q$ is the 
sensitivity of the scoring function, i.e., the maximum change in assigned scores 
possible for two neighbouring datasets.

Both above mechanisms assume that $\Delta 
f$ and $\Delta q$ are known beforehand. 
%In applications where these values 
%cannot be determined, 
The \emph{sample-and-aggregate 
framework}~\cite{nissim2007} drops this assumption
%allows the privatization of functions, for which 
%these values are unknown 
by sampling subsets of the input set and evaluating the given function per 
sample. The obtained results are then combined using a known 
differentially private aggregator. If function $f$ can be approximated 
well on small sub-samples, then the results per sample are close to $f(X)$. By 
aggregating these approximated results using a differentially private mean, 
i.e., by computing the mean and adding noise 
drawn from a Laplacian calibrated with $\Delta(mean)$,  one achieves a 
differentially private result for $f(X)$. 
%even though the sensitivities of the 
%function are unknown.
%\todo{Stephan: If we have space, I think ann example would 
%improve the explanation of the sample-and-aggregate framework. Right now, I 
%find it hard to follow.}

	\section{A Framework for Privacy-Aware PPIs}\vspace{-.5em}
	\label{sec:privacymodel}
	% !TeX spellcheck = en_GB
% !TeX root = ../main.tex

In this section, we introduce a generic framework for the evaluation of 
privacy-protected PPIs, thereby addressing the research question raised 
above. Specifically, we discuss design decisions, as well as the underlying 
assumptions and limitations of the framework.

\begin{figure}[h!]
  \vspace{-1em}
  \centering
  \includegraphics[scale=0.2]{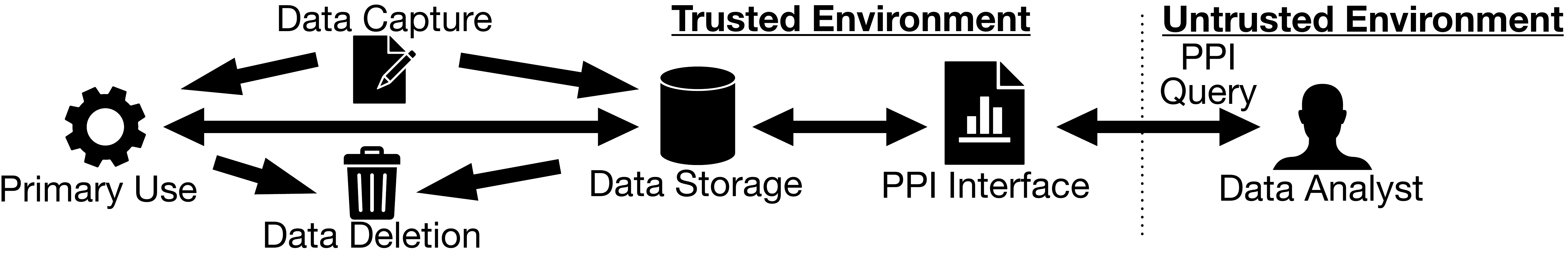}
  \vspace{-.5em}
  \caption{Overview of the framework for privacy-aware PPIs.}
  \label{fig:protection_model}
  \vspace{-2.25em}
\end{figure}

The evaluation of PPIs is usually conducted on event data recorded and 
administered by the process owner. As such, we consider a centralized model and 
assume that the entity collecting and persistently storing the event data is 
trustworthy. However, as illustrated in 
\autoref{fig:protection_model}, the actual information demand regarding the 
PPIs is external to this environment. Following existing models for the flow of 
data in the analysis of information systems, the evaluation 
of PPIs is conducted in an untrusted 
environment~\cite{daquisto2015,MannhardtPO18}.

Considering the handling of event data in the trusted environment in more 
detail, the following phases are distinguished. First, the \emph{Data Capture} 
phase concerns the collection of process-related event data, i.e., whenever an 
activity has been executed, a respective event is created. Subsequently, the 
phase of \emph{Primary Use} represents that the captured event data is 
exploited for the purposes for which it was recorded, which is commonly the 
proper execution of an individual instance of the process. For example, the 
recorded events may be used to invoke services, trigger notifications, or 
schedule tasks for knowledge workers. At the same time, the event data is made 
persistent, which is modelled 
as a \emph{Data Storage} phase. The persisted event data may then be used for 
\emph{Secondary Use}, such as process improvement initiatives conducted by 
process analysts. Eventually, the 
event data may be deleted from the persistent storage, in a \emph{Data 
Deletion} 
phase. All phases, except the \emph{Secondary Use}, are conducted within the 
realms of the trusted environment. The \emph{Secondary Use} is part of the 
untrusted environment, since the data was recorded without having any consent 
on their use for these applications.

Unlike common primary use of process-related event data, process improvement in 
general, and the computation of PPIs in particular, aim at generalizing the 
observations made for individual process instances in some aggregated measures. 
%PPIs are a secondary use, with the aim to calculate aggregated measures on the 
%basis of many individual process instances.
Thus, in these contexts, the privacy of an involved individual would be 
compromised, if their contribution to the published aggregate would be revealed 
to the process analyst.
To enable such secondary use without compromising an individual's privacy, we 
need to prevent a process analyst to 
assess the impact of a single process instance on the aggregated result. Hence, 
we consider each trace and the information inferred from it as 
sensitive information. For example, when a PPI is based on the mean sojourn 
time of all process instances, we aim to protect the specific sojourn time of 
each instance.

To achieve this protection, 
%While all phases, that lie within the trusted environment come with full 
%access to any data captured and produced during system execution, such 
any access to the event data from the untrusted environment must be 
restricted. Therefore, we propose to design an interface for the evaluation of 
PPIs, thereby realizing an explicit privacy checkpoint. The interface receives 
PPI queries stated in PPINOT syntax and answers them while ensuring 
differential privacy. To this end, the interface fetches relevant event data 
from the persistent storage based on the scope of the PPI query, calculates the 
result, and adds noise to the result, before releasing the result to the process 
analyst. Any such release reduces a privacy budget, which is chosen based on 
the desired strength of the privacy guarantee to implement. 
Since the noise added to the result is calibrated based on the specifics of a 
PPINOT query and the event data retrieved for evaluating it, 
we ensure $\epsilon$-differential privacy. 

By the above, we achieve \emph{plausible deniability}: An analyst 
cannot distinguish between 
query results that contain a particular process instance and those that do not.
%As PPIs are evaluated on disjoint subsets of the recorded data with different 
%time frame specifications for each PPI, we use a privacy budget for each 
%recorded process instance. Whenever a PPI query accesses the data of a process 
%instance, the budget of the process instance is reduced accordingly?????

While access to the event data is restricted, our framework assumes that an 
analyst has access to models of the respective processes in order to 
specify the PPIs. 
%While it is obvious, that any data analyst, querying the interface must not 
%have access to the stored raw data, the access to the underlying process 
%models 
%is necessary, as the PPIs are defined on the basis of these models. %Likewise, 
%%the interface needs to be in possession of a mapping from the activities of 
%%the 
%%defined process models to the recorded events in the process 
%%logs\todo{Stephan: 
%%Diesen Satz verstehe  ich nicht}.
Another assumption of our framework is that, for a given time scope, an upper 
bound for the appearances of an individual in the recorded process instances is 
known. An individual appearing in $n$ process instances dilutes 
$\epsilon$-differential privacy by at most $e^{\epsilon n}$. Knowing an upper 
bound for $n$, however, enables mitigation of this effect by changing the 
privacy parameter $\epsilon$ accordingly. For our example in 
\autoref{fig:running}, we would need to know the maximal number of claims that 
can be handled by a knowledge worker within a single month.
%Another assumption of our framework is that each individual corresponds to 
%exactly one recorded process instance in a given time scope. Otherwise, the 
%impact of an individual on the result of a PPI may be larger than what will be 
%accounted for in the privacy guarantee. Assuming that an individual can be 
%part 
%of $n$ instances, the privacy guarantee is diluted by at most $e^{\epsilon 
%n}$. 
%%Nonetheless, this risk is minimized, due to the fact, that the recorded data 
%%is evaluated based on the query scope.  
%Hence, if an upper bound for the appearances of an individual is known, this 
%effect is mitigated by changing the privacy parameter $\epsilon$ accordingly. 
Lastly, we acknowledge that, while we focus on the evaluation of 
PPIs, 
%since no methodology alone serves as a catch-all silver bullet to 
further privacy threats in the trusted environment require additional 
protection mechanisms\cite{daquisto2015}. 
%Examples for such 
%means include 
%%applies additional protection measures, e.g. by 
%access control, data partitioning, and data encryption. 
%, and data channel 
%encryption.

	\section{Release Mechanisms for Privacy-Aware PPIs}\vspace{-.5em}
	\label{sec:evaluationtechniques}
	% !TeX spellcheck = en_GB
% !TeX root = ../main.tex

In this section, we instantiate the above framework and introduce a specific 
realization of the interface for the evaluation of PPIs. 
%The interface achieves privacy protection by adding $\epsilon$-differentially 
%private noise to query results. 
We first show how the interface leverages the compositional structure 
of PPIs defined based on the PPINOT meta-model in 
\autoref{subSec:CompositionTreePrivacy}. We then provide a set of 
$\epsilon$-differentially private release mechanisms in 
\autoref{subSec:releaseMechanisms}.

\subsection{Using Function Composition Trees for Privacy Protection}
\label{subSec:CompositionTreePrivacy}

Our idea is to exploit the compositional nature of PPIs defined in the PPINOT 
meta-model for privacy protection. Instead of adding noise to the final query 
result, we introduce noise, with smaller magnitude, at the inner functions of a 
PPI. Such a compositional approach still guarantees $\epsilon$-differential 
privacy of the result. At the same time, it enables us to minimize the 
overall introduced error. Hence, data utility is preserved to a 
higher degree, which leads to more useful process analysis, under the same 
privacy guarantees. 

We aim to protect the privacy of individuals, of whom personal data is
materialized in a trace. Hence, the results of single-instance measures (base 
or derived measures) shall be protected. However, common PPIs assess the 
general performance of process execution by aggregating these results in 
multi-instance measures (aggregation or derived measures), so that guarantees 
in terms of differential privacy may be given for these measures. This 
raises the question of selecting a subset of the multi-instance measures for 
privatization. On the one hand, this selection shall ensure that the results of 
\emph{all} aforementioned single-instance measures are protected. On the other hand, the 
selection shall be \emph{minimal} to keep the introduced noise to the 
absolutely necessary magnitude. 

%The first property ensures, that knowledge extracted from each leaf node is 
%privatized in at least function. The second function serves as a minimality 
%constraint, ensuring, that the set of functions is minimal. This minimizes the 
%introduced noise magnitude. 

%Hence, our 
%focus for privatization is on 
%% our of Differential 
%%privacy applies to aggregations of multiple inputs. Therefore,
%% PPINOT single-instance measures are no valid candidates for privatization. 
%%Thus, only 
%aggregation measures and multi-instance derived measures. 
% may serve as 
% applicable functions for privatization. 
%Our goal is to protect sensitive information of each participant, i.e. the raw 
%data extracted from the recorded process instances by single-instance 
%measures. 
%Therefore the set of selected measures needs to privatize the information 
%extracted by all single-instance measures that retrieve data from the recorded 
%process data. 
%These measures transitively depend on the results of single-instance measures, 
%which correspond to leaf nodes in the function composition tree. 

We capture the above intuition with the notion of an admissible set of measures 
of a PPI. Let $(F,\rho)$ be the function composition tree of a PPI, with $F$ as 
the set of measures and $\rho: F\rightarrow 2^F$ as the function assigning 
child measures to measures. With $\rho^*$ as the transitive closure of $\rho$, 
a set of measures $F'\subseteq F$ is admissible, if it:
\begin{compactitem}
  \item contains only multi-instance measures: $f\in F'$ implies that 
  $f\in \dom(\rho)$;
  \item covers all trace-based measures: $\forall \ f\in (F\setminus 
  \dom(\rho)): \exists \ f'\in F': f\in \rho^*(f')$; 
  \item is minimal: $\forall \ F''\subset F': \exists \ f\in (F\setminus 
  \dom(\rho)): \forall \ f''\in F'': f\notin \rho^*(f'')$.
\end{compactitem}
% be the set of 
%measures of a PPI, $\Gamma \subset F$ 
%
%Let $F_1,\ldots, F_n \subset F$ be the set of candidate 
%functions for privatization. Let $\Gamma \subset F$ be the set of leaves of 
%the 
%tree.
%Then a given set $F_i$  is admissible for privatization if the following 
%properties hold:
%\begin{align*}
%\forall l\in \Gamma: \exists f\in F_i: f\in \tau(l)\\
%\nexists F_j \subset F: F_j\subset F_i, F_j \text{is admissible}\\
%\end{align*} 

\smallskip
\noindent
The first condition of an admissable set applies, as differential privacy may only be used for the aggregation of multiple inputs, thus single-instance measures cannot be privatized with the given privacy framework. The second condition ensures, that the selected set of functions privatizes all single-instance derived or base measures, that directly access trace information. Finally, the third condition ensures, that only the minimum amount of noise to achieve $\epsilon$-differential privacy is added onto the intermediate results.

The function composition tree in \autoref{fig:function_composition_tree} has 
two sets of admissible measures, $\{r\}$ and $\{sum_1,sum_2\}$, which both 
cover the single-instance measures $count$(`RejC') and $count$(`RecC'). In 
contrast, the set 
$\{sum_1\}$ is not admissible, as $count$(`RecC') is not covered (second 
constraint). Likewise, selecting both base measures or $\{r,sum_1,sum_2\}$ is 
not admissible, as this would violate the first and third constraint, 
respectively.

Once a set of admissible measures is selected, the evaluation of the PPI is 
adapted by incorporating a release mechanism, as defined next, for the chosen 
measures. 
%Next, we 
%turn to the definition of these release mechanisms.

\subsection{Release Mechanisms for Multi-Instance Measures}
\label{subSec:releaseMechanisms}

%We propose three release mechanisms for the privatization of 
%aggregation measures, followed by one mechanism for derived multi-instance 
%measures. 

The design of a release mechanism for a specific multi-instance measure is 
influenced by (i) the ability 
to assess the domain of input values over which the measure is evaluated, and 
(ii) the ability to assess the sensitivity of the measure. As for 
the first aspect, the PPI interface of our framework, see 
\autoref{sec:privacymodel}, can rely on an estimation of the respective domain. 
Here, a simple estimation is based on the minimal and maximal values, 
$\underline{X}$ and $\overline{X}$, of the 
dataset $X$ used as input for the measure (i.e., the result of the child 
measures). The bounds may be extended by constant offsets to account 
for the fact that the dataset $X$ is merely a sample of an unknown 
domain. The sensitivity of the measure, in turn, depends on the semantics of 
the measure. While for the aggregation functions of PPINOT, this sensitivity 
may be estimated, it is unknown in the general case of derived measures. 

Against this background, this section first introduces three release mechanisms 
for aggregation measures: an instantiation of the Laplace mechanism; an 
interval-based mechanism based on the traditional 
exponential mechanism; and a threshold-sensitive mechanism that extends the 
interval-based one to preserve the significance of a measure related to a 
threshold. Finally, we discuss how derived measures, in the absence of an 
estimate of their sensitivity, can be privatized using a sample-and-aggregate 
strategy.

% before discussing how to handle derived multi-instance measures. 
%All release 
%, we have to assume that the effective 
%range of input values of the dataset $X$, over which a multi-instance measure 
%is computed, is known in advance.
%%All of the presented measures assume, that the effective range of 
%%input values in the input set $X$ are known in advance. 
%As a querying process 
%analyst must not have access to the raw input data, simply determining these 
%values is not possible, without risking privacy intrusions. Yet, these values 
%can be determined based on the selected input set $X$, either by using the 
%minimum and maximum value as boundary or by extending these ranges by a 
%constant factor. The first method achieves smaller noise magnitudes, while the 
%second method resembles, that $X$ is merely a sample of a larger unknown input 
%space. In the following, we assume that the input ranges are determined on 
%either one of these two methods, or any other method, that includes all values 
%of $X$.

%\subsubsection{Mechanisms for Aggregation Measures}

%For aggregated measures we propose the application of the well-established 
%Laplace mechanism and the Exponential mechanism.

\medskip
\noindent
\textbf{{Laplace Mechanism for Aggregation Measures.}}
Privatization of an aggregation measure 
%Since the aggregation functions have known and well-defined behaviour, we can 
%apply the Laplace mechanism for privatization, i.e. the 
can be based on the addition of Laplace noise to the actual result. 
As mentioned, this requires to estimate the sensitivity $\Delta f$ of the given 
aggregation function, i.e., the maximal impact any element $x\in X$ may have on 
$f(X)$. For the aggregation functions of the PPINOT meta-model, the sensitivity 
is derived as $\Delta(min)=\Delta(max)=|\overline{X}-\underline{X}|$,  
$\Delta(sum)=\overline{X}$ and 
$\Delta(mean)={|\overline{X}-\underline{X}|}/{|X|}$.
%\todo{give worst-case sets or not? Matthias - what is that? Martin: The sets 
%from which the worst case changes are derived} 
Based thereon, noise from a Laplacian (with parameters $\mu=0$ and $b={\Delta 
f}/{\epsilon}$, see \autoref{sec:differential_privacy}) is added to $f(X)$. 

Since the sensitivity $\Delta f$ directly influences the magnitude of added 
noise, for $mean$ measures, this mechanism potentially leaks information about 
the number $|X|$ of process instances (and hence, individuals) within the 
given scope. An adversary may conclude on the difference 
$|\overline{X} - \underline{X}|$ based on the magnitude of noise from another 
PPI incorporating a $min$ or $max$ measure and, based thereon, derive $|X|$ 
from the magnitude of noise in a PPI with a $mean$ measure. However, in 
practice, $|X|$ may be revealed explicitly to enable a 
process analyst to assess the statistical reliability of the PPI result. 

%seem a disadvantage, yet in 
%practice the treatment of this knowledge as public knowledge is desirable. 
%Even 
%without privatization, the goodness of a result relates directly to the number 
%of instances it was calculated upon, so this would allow a process analyst to 
%assess the statistical reliability of the result.

%While leakage of $|X|$ may seem a disadvantage, yet in 
%practice the treatment of this knowledge as public knowledge is desirable. 
%Even 
%without privatization, the goodness of a result relates directly to the number 
%of instances it was calculated upon, so this would allow a process analyst to 
%assess the statistical reliability of the result.

\medskip
\noindent
\textbf{Interval-based Mechanism for Aggregation Measures.}
The drawback of the Laplace mechanism to privatize aggregation measures 
is the inherently high sensitivity, which scales linearly with the domain of 
input values. Our idea, therefore, is to group similar result values into 
intervals and score them using the exponential mechanism. This way, we obtain a 
release mechanism with a score function sensitivity $\Delta q=1$, which 
ultimately leads to a smaller magnitude of noise for large domains of input 
values. 

To realize this idea, our interval-based release mechanisms consists of three 
phases: 
\begin{compactenum}[(1)]
\item \textit{Interval creation:} We partition the domain and the range of the 
aggregation function into a set of intervals.
\item \textit{Interval probability construction:} Scores are assigned to these 
intervals, which are then converted to result probabilities.
\item \textit{Result sampling:} Using these probabilities, an interval is 
chosen as the output interval, from which the result value is sampled. 
\end{compactenum}
%\textit{interval creation}, \textit{interval probability creation}, and 
%\textit{result sampling}. During \textit{interval creation}, we 
%partition the domain and the range of the aggregation function into a set of 
%intervals. In the \textit{interval probability creation} phase, scores are 
%assigned to these intervals, which are then converted to result 
%probabilities. Lastly, using these probabilities, an interval is chosen as the 
%output interval and the final output value is chosen in the \textit{result 
%sampling} phase. 
%In the following, we present the three phases and how these 
%are instantiated for each of the four aggregation functions.
The \textit{interval creation} is based on the range of the 
aggregation function, given as $range(f(X))=(\underline{X},\overline{X})$ for 
$f\in \{min,max,mean\}$ and $range(f(X))=(\underline{X}\cdot |X|,\overline{X} 
\cdot |X|)$ for $f=sum$. This range is split into non-overlapping 
intervals $I=\{I_0,\ldots,I_n\}$,  with $I_0 \cap \ldots \cap 
I_n=\emptyset$ and $I_0 \cup \ldots \cup I_n=range(f(X))$.
%splits up the range of the aggregation function possible output values 
%$R$ is converted into a set of non-overlapping intervals, i.e. 
%$I=\{I_0,I_1,...,I_n\}$ with $I_0 \cap I_1 \cap ... \cap I_n=\emptyset$ and 
%$I_0 \cup I_1 \cup ... \cup I_n=R$. For each of the four aggregation measures, 
%the output range $R$ is known beforehand and can be computed based on the 
%input 
%set $X$. 
%These ranges of $R$ are given as 
%$R_{min,max,mean}=(\underline{X},\overline{X})$ and 
%$R_{sum}=(\underline{X}\cdot |X|,\overline{X} \cdot |X|)$.
Let $\tau(x_i, x_j)={(x_i + x_j)}/{2}$ be the mean of $x_i, x_j$ 
%let $f(X)$ be the true result of function f and let 
and let $I_f$ be the interval containing the result value, i.e. %for which 
$f(X)\in I_f$. For $mean$ and $sum$, the range of $f(X)$ is divided into evenly 
spaced intervals of size $\Delta f$, so that 
$f(X)=\tau(\underline{I_f},\overline{I_f})$ is the mean of its 
containing interval.   
%i.e. $I={(\underline{R},f(X)-m \Delta f),...,(f(X)-1.5 \Delta f,(f(X)-\frac{\Delta f}{2}),(f(X)-\frac{\Delta f}{2},(f(X)+\frac{\Delta f}{2}),(f(X)+\frac{\Delta f}{2},f(X)+1.5\Delta f),...,}$
For $min$ and $max$, the range of $f(X)$ is divided into $n$ intervals of 
different size, for which the boundaries are the means of 
neighbouring values $\tau(x_i, x_{i+1})$ with $x_i,x_{i+1}\in X$.

\begin{figure}[!t]
	\centering
	\includegraphics[scale=0.47]{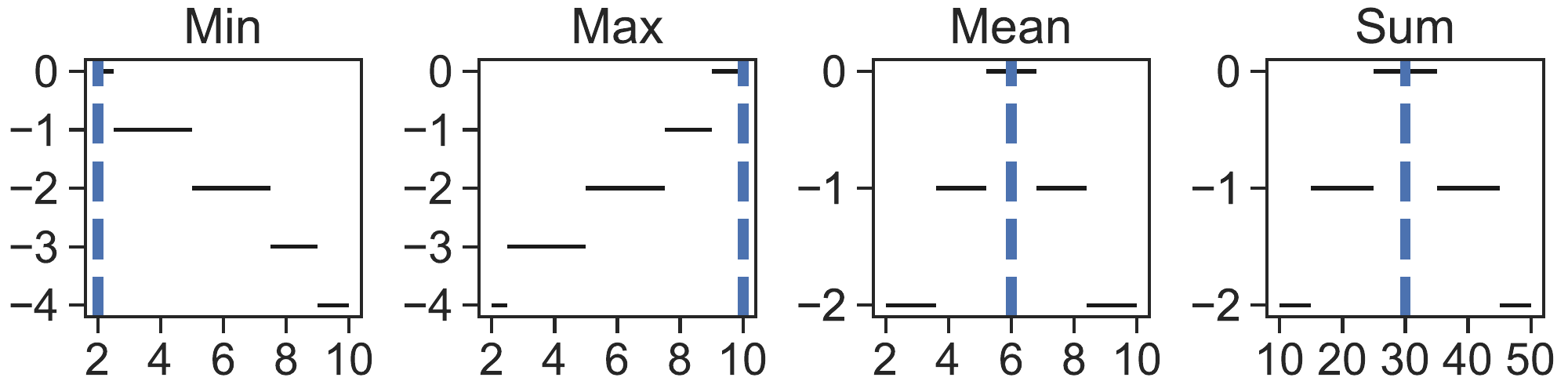}
	\vspace{-.5em}
	\caption{Intervals and scores for the aggregation functions 
	for dataset $X=\{2, 3, 7, 8, 10\}$.}
	\label{fig:interval_scores}
	\vspace{-1em}
\end{figure}

\autoref{fig:interval_scores} exemplifies the intervals for a dataset $X=\{2, 
3, 7, 8, 10\}$. For $min$ and $max$, the interval boundaries are $2.5$, $5$, 
$7.5$, and $9$. For $mean$ and $sum$, the intervals have size $\Delta f=1.6$ 
and $\Delta f=10$, and are centred around $mean(X)=6$ and $sum(X)=30$.

%$I={(\underline{R},\underline{R}+\Delta f),(\underline{R}+ \Delta 
%f,\underline{R}+2\cdot \Delta f), ...,(\underline{R}+ m\cdot \Delta 
%f,\overline{R})}$, i.e. the output space is divided into m evenly-sized 
%intervals of size $\Delta f$. For minimum and maximum functions 
%$I={(\underline{R},\tau(x_1, x_2)),(\tau(x_1, x_2), \tau(x_2, x_3)), ..., 
%,(\tau(x_{n-1},x_n), \overline{R})}$. Here, R is divided into n intervals of 
%different size, whose boundaries correspond to the midpoints between 
%neighbouring values $x_i, x_{i+1}$. For the input set $X={x_1, x_2, x_3, x_4, 
%x_5, x_6}$ the created intervals are illustrated in 
%\autoref{fig:interval_creation}. These are 
%$I_{mean}=\{(1,2.6),(2.6,4.2),(4.2,5.8),(5.8,7.4),(7.4,9.0)\}$,
%$I_{sum}=\{(5, 9),(9, 18),(18, 27),(27, 36),(36, 45)\}$ and $I_{min}=I_{max}=\{(1,2),(2,4),(4,6),(6,8),(8,9)\}$.

The \textit{interval probability construction} relies on a scoring function 
that assigns higher scores to intervals that are closer 
to the interval containing $f(X)$. 
%ranking  That  that is an 
%interval is ranked higher, the closer it is to the interval containing the 
%true 
%function result $f(X)$.
Let $I_1,\ldots, I_n$ be the intervals in the order induced by $\leq$ over 
their boundaries, and let $1\leq k\leq n$ be the index of interval $I_f$ 
containing the result value. 
%${1\leq 
%j\leq n}$ be the interval, for which $\underline{I_j}\leq f(X) \leq 
%\overline{I_j}$. 
Then, the score for each interval $I_{i}$ is defined as $q(i)=-|k-i|$, as 
illustrated in \autoref{fig:interval_scores} for the example. Here, intervals, that lie closer to $f(X)$, denoted by the blue dashed lines, are scored higher, than those further away.
Since each interval $I_i$ corresponds to a set of potential result values, we 
incorporate the size of this set in the probability computation. Hence, the 
probability for $I_i$ is defined as:
\begin{align*}
P(I_i)=\frac{|I_i| \cdot e^{(\nicefrac{\epsilon\cdot q(i)}{2\cdot \Delta 
q})}}{\Sigma_{1\leq j\leq n}|I_j| \cdot e^{(\nicefrac{\epsilon\cdot 
q(j)}{2\cdot \Delta q})}}
\end{align*}

\textit{Result sampling} chooses one interval based on their 
probabilities. From this interval one specific value is drawn based on a 
uniform distribution over all interval values.
%Since the expected value of a random variable drawn from a uniform distribution of interval $I$ is given as $\frac{\underline{I}-\overline{I}}{2}$, we can achieve results, that lie closer to $f(X)$, should the interval containing $f(X)$ be sampled. By enforcing, that $f(X)$ coincides with the expected value of its corresponding interval, the accuracy of the mechanism may be improved. For mean and sum functions this can be achieved by shifting all intervals accordingly, while for minimum and maximum functions the doubling. To accommodate this shift, the size of the topmost or lowermost interval is reduced, while a new interval needs to be added on the opposite side of the output range $R$.
%For minimum and maximum functions, $f(X)$ coincides with $\underline{X}$ and $\overline{X}$ respectively. By setting $\underline{I}=\underline{I}-|\overline{I}-\underline{I}|$ for minimum functions and $\overline{I}=\overline{I}+|\overline{I}-\underline{I}|$ for maximum functions, $f(X)$ becomes the expected value of it's interval.

\medskip
\noindent
\textbf{{Threshold-Sensitive Mechanism for Aggregation Measures.}}
The interval-based mechanism is problematic, if a PPI is tested against a 
threshold, as often done in practice. Consider the 
dataset $X$ and assume that the $sum$ function is the root of a PPI's function 
composition tree, i.e., $f(X)=30$ as shown in \autoref{fig:interval_scores}. 
Assume that it is important whether the PPI is less or equal than $30$. 
Then, adding noise may change the actual 
interpretation of the PPI, since the release mechanism will sometimes publish 
values larger than 30. 
%goal is reached, 
%yet the addition of noise may result in a final result, which signifies, that 
%this target value has not been reached, all tough the opposite would be true. 

To mitigate this effect, we present a threshold-sensitive release mechanism 
that extends the previous mechanism in terms of \textit{interval 
  creation} 
and \textit{interval probability construction}. Let $\chi$ be a Boolean 
function formalizing a threshold, e.g., $\chi(x)= x \leq 30$. Then, 
the Boolean predicate $\phi(x,f(X),\chi) \Leftrightarrow \chi(x) \equiv 
\chi(f(X))$ describes, whether the possible result value $x\in range(f(X))$ 
leads to the same outcome of $\chi$ as the true result $f(X)$. For 
our example, 
%the sum function and $\chi(x)= x \leq 30$, 
$\phi(20, 30,\chi)$ holds true ($20\leq 30$ and $30\leq30$), whereas $\phi(40, 
30,\chi)$ is false ($40\nleq 30$, but $30\leq30$). 

\begin{wrapfigure}{r}{0.4\textwidth}
  \vspace{-.2em}
  \centering
  \includegraphics[scale=0.5]{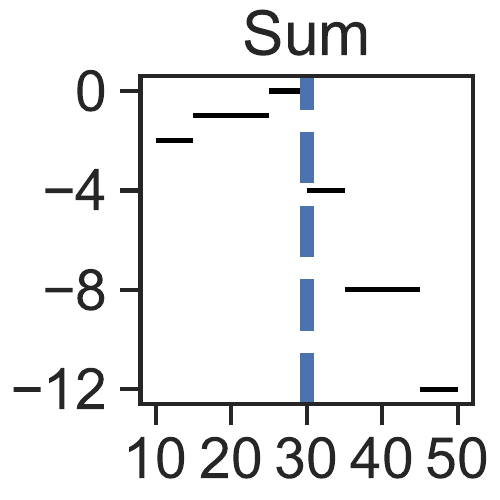}
  \vspace{-.5em}
  \caption{Adapted intervals and scores.}
  %\caption{Scores assigned to the intervals created in 
  %  \autoref{fig:interval_scores} for the sum function, in accordance to 
  %  $\phi(x):x\leq 30$}
  \label{fig:preserving_intervals}
  \vspace{-2.25em}
\end{wrapfigure}

Using this predicate, we adapt the intervals $I=I_1,\ldots,I_n$ obtained during 
\textit{interval creation}, so that interval boundaries coincide with changes 
in $\phi$. 
%Specifically, we augment the interval-based mechanism as follows
Let $B(\phi)$ be the boundary values of $\phi$, i.e., the values $x\in 
range(f(X))$ with $\lim_{y<x, y\rightarrow x}\phi(y,f(X),\chi)\neq 
\lim_{y>x, y\rightarrow x}\phi(y,f(X),\chi)$. For our example, we arrive at 
$B(\phi)=\{30\}$. Based thereon, we split each interval $I_i$ containing a 
boundary value $b\in B(\phi)$ into two new intervals 
$(\underline{I_i},b),(b,\overline{I_i})$. Hence, each interval contains only 
values that share the outcome of the Boolean function $\chi$. 
In our example, the interval $(25,35)$ is split into $(25,30)$ and $(30,35)$, 
as shown in \autoref{fig:preserving_intervals}.

%\begin{figure}[!t]
%	\centering
%	\includegraphics[scale=0.6]{fig/interval_scores_preserving.pdf}
%	\caption{Scores assigned to the intervals created in 
%\autoref{fig:interval_scores} for the sum function, in accordance to 
%$\phi(x):x\leq 30$}
%	\label{fig:preserving_intervals}
%\end{figure}

Finally, the scoring function used for \textit{interval probability 
construction} is adapted. Let $d(i)$ be the minimal inter-interval-distance of 
interval $I_i$ to any other interval $I_j$ with $\phi(x,f(X),\chi)\neq 
\phi(y,f(X),\chi)$ for all $\underline{I_i}\leq x\leq \overline{I_i}$
and $\underline{I_j}\leq y\leq \overline{I_j}$. 
As before, let $k$ be the index of interval $I_f$ containing the result value. 
Then, scores assigned to intervals %$I_i$ 
that preserve the outcome of the 
Boolean function $\chi$
%(i.e., $\phi(x,f(X),\chi)$ is true for all $ \underline{I_i}\leq x\leq 
%\overline{I_i}$) 
remain unchanged. For all other 
intervals $I_i$, the score is reduced by $\xi\cdot d(i)$, i.e., by the distance 
to the closest interval preserving the outcome multiplied by a falloff factor 
$\xi\in\mathbb{N}$. The adapted scoring function is defined as:
\[
q(i)=\begin{cases}
-|k-i| &\text{ if }\phi(x,f(X),\chi) \text{ for all } \underline{I_i}\leq x\leq 
\overline{I_i},\\
-|k-i|-\xi \cdot d(i) &\text{ otherwise.}
\end{cases}\]
\autoref{fig:preserving_intervals} illustrates the adapted scores for our 
running example, using $\xi=3$. The scores of the 
right-most three intervals are reduced, as all of their values lead to a 
different outcome compared to the true result, $f(X)=30$, when testing against 
$\chi(x)= x \leq 30$. 
We obtain $d(4)=1, d(5)=2$, and $d(6)=3$ for those intervals, given that the 
third interval $(25, 
30)$ is the closest one retaining $\phi$ to any of those three. Thus, we arrive 
at $q(4)=-4, q(5)=-8$, and $q(6)=-12$.
As the largest possible change in scores assigned to a possible result value in 
neighbouring input sets is never larger than $\xi$ and as the interval sizes 
are determined based on $\Delta f$, we conclude that $\Delta q=\xi$.

\medskip
\noindent
\textbf{Sample-and-aggregate Mechanism for Derived Measures.}
Since the sensitivity of a derived multi-instance measures is unknown in the 
general case, the above mechanisms are not applicable. 
% of the various aggregation 
%functions are known beforehand they lend themself to the application of the 
%already presented mechanism. Yet, for derived multi-instance measures, this 
%information is usually not present, as these can take an arbitrary form. 
However, many derived measures may be approximated using small samples, since 
their range is often independent of the domain of their input values. Functions 
that compute a normalized result are an example of this class of 
measures. For instance, the derived measure that denotes the root of the 
function composition tree of the example in 
\autoref{fig:function_composition_tree} yields a percentage, i.e., it is 
normalized to 0\% to 100\%. For such measures, the 
sample-and-aggregate-framework mentioned in \autoref{sec:differential_privacy} 
may be instantiated. 
%While the sensitivity of a function may be hard to prove beforehand, the 
%output range 
%may be independent on the input values, and as such known beforehand. 
%Functions, that normalize their result, are an example of this class of 
%functions. 
That is, the actual result $f(X)$ is computed on $n$ partitions of $X$. The 
obtained results per sample are then aggregated using a differentially private 
mean function to achieve privatization of the derived measure.

	\section{Experimental Evaluation}\vspace{-.5em}
	\label{sec:evaluation}
	% !TeX spellcheck = en_GB
% !TeX root = ../main.tex

To assess the feasability and utility of the proposed approach, 
we realized the PPI interface on top of an existing 
PPINOT implementation.\footnote{
	\url{https://mvnrepository.com/artifact/es.us.isa.ppinot/ppinot-model}}
We conducted 
controlled experiments using synthetic data 
(\autoref{subSec:controlledExperiments}) and a
case study with the \textit{Sepsis Cases} log 
(\autoref{subSec:caseStudy}). The latter compares the proposed tree-based 
privatization with the direct evaluation of PPIs on logs that have been 
anonymized with the PRIPEL framework~\cite{fahrenkrog-petersen2020} beforehand.
 Our implementation and evaluation scripts are publicly 
 available.\footnote{\url{github.com/Martin-Bauer/privacy-aware-ppinot}}
%In the following, we first report on the results of the evaluation using 
%synthetic data (\autoref{subSec:controlledExperiments}), followed by the 
%results of our case study (\autoref{subSec:caseStudy}).

\subsection{Controlled Experiments}
\label{subSec:controlledExperiments}
In a first series of experiments, we assessed the impact of different 
properties of the dataset $X$ used as input. 
Specifically, we consider the impact of the estimation of the domain of input 
values, its size and underlying value distribution, and the privacy parameter 
$\epsilon$. We sampled sets of 10, 50, 100, and 200 random values from a 
Gaussian distribution, a Pareto distribution, and a Poisson distribution. We 
chose these distributions, as they are often observed in event data recorded by 
business processes. %To account for the randomness of our mechanisms, 
We performed 200 runs per experiment.
Unless noted otherwise, the input domain is estimated using the minimal and 
maximal element of $X$, the dataset comprises 200 values drawn from a Gaussian 
distribution, and the privacy parameter is set as $\epsilon=0.1$.
%For our evaluation we use $\epsilon=0.1$, input range estimation using the 
%smallest and largest element of $X$ and the Gaussian distribution using 200 
%values, and change the property under investigation, while keeping the others 
%fixed.

\begin{figure}[!t]
	\centering
	\includegraphics[scale=0.38]{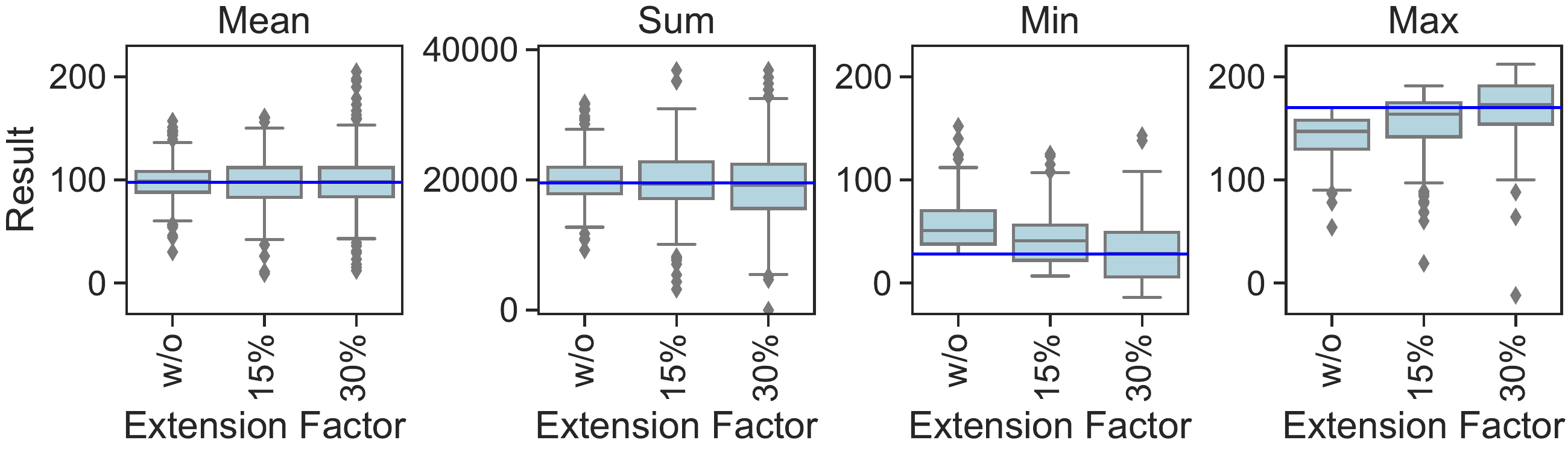}
		\vspace{-.8em}
	\caption{Impact of the input boundary estimation on the results.}
	\label{fig:boundary_estimation}
	\vspace{-2.25em}
\end{figure}

\medskip
\noindent
\textbf{{Input Boundary Estimation.}}
First, we compare the boundary estimation using the minimal and maximal 
elements in $X$ with extensions of these boundaries by $15\%$ and $30\%$ at 
either boundary. The results for the interval-based mechanism, see 
\autoref{fig:boundary_estimation}, show that an extension of the domain
increases the introduced magnitude of noise for all functions due to an 
increase in sensitivity. These observations are 
confirmed for the Laplace mechanism. Yet, for 
$min$ and $max$, there is a shift of the expected result 
towards the true result $f(X)$ (denoted by the blue line). The reason is that, 
without the extension, $f(X)$ coincides with boundary values of $X$. 
The extension increases the size of the interval containing $f(X)$, which 
increases the probability of this interval to be chosen.

\begin{wrapfigure}{r}{0.45\textwidth}
	\vspace{-1.5em}
	\centering
	\includegraphics[scale=0.38]{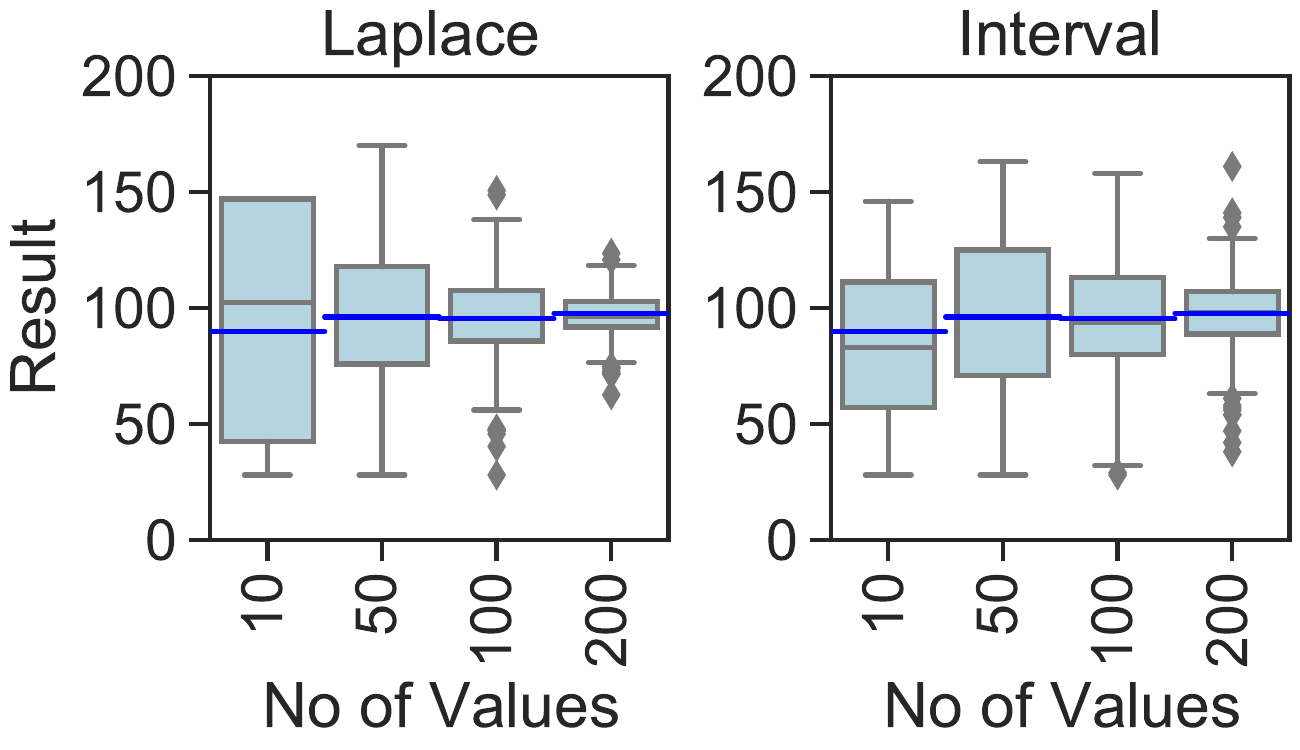}
%\vspace{-.5em}
\caption{Impact of the cardinality of the dataset $X$ on the 
	results for $mean$.}
\label{fig:noOfValues_mean}
	\vspace{-2.25em}
\end{wrapfigure}

\medskip
\noindent
\textbf{{Input Size and Distribution.}}
For the Laplace and interval-based mechanisms, we identify a dependency of 
$\Delta f$ on the input size for $mean$ functions. This dependency coincides 
with smaller noise magnitudes for larger input sizes, as illustrated in 
\autoref{fig:noOfValues_mean}.

%\begin{figure}[!t]
%	\centering
%	\includegraphics[scale=0.4]{fig/noOfValues_mean.pdf}
%	\vspace{-.5em}
%	\caption{Impact of the cardinality of the dataset $X$ on the 
%	results for $mean$.}
%	\label{fig:noOfValues_mean}
%	\vspace{-1em}
%\end{figure}

These trends were confirmed for the interval-based mechanism for $min$ and 
$max$. Here, the increased number of intervals and a more 
fine-grained differentiation between result values leads to higher utility, 
i.e., the expected result is close to the actual one.  
Yet, the trends are only visible for distributions with small 
inter-value distances, such as the Gaussian. For the Pareto- 
and Poisson-distributions, there was a significant reduction in 
utility for larger inputs using $max$. These 
distributions preserve most of their 
probability mass on the smaller values, 
%Larger values are therefore more 
%scarce, 
%which 
This inadvertently 
results in the creation of disproportionally large intervals and the same 
output probability for large portions of the output space.

\begin{figure}[!t]
	\begin{subfigure}[b]{0.5\textwidth}
	\centering
		\includegraphics[scale=0.38]{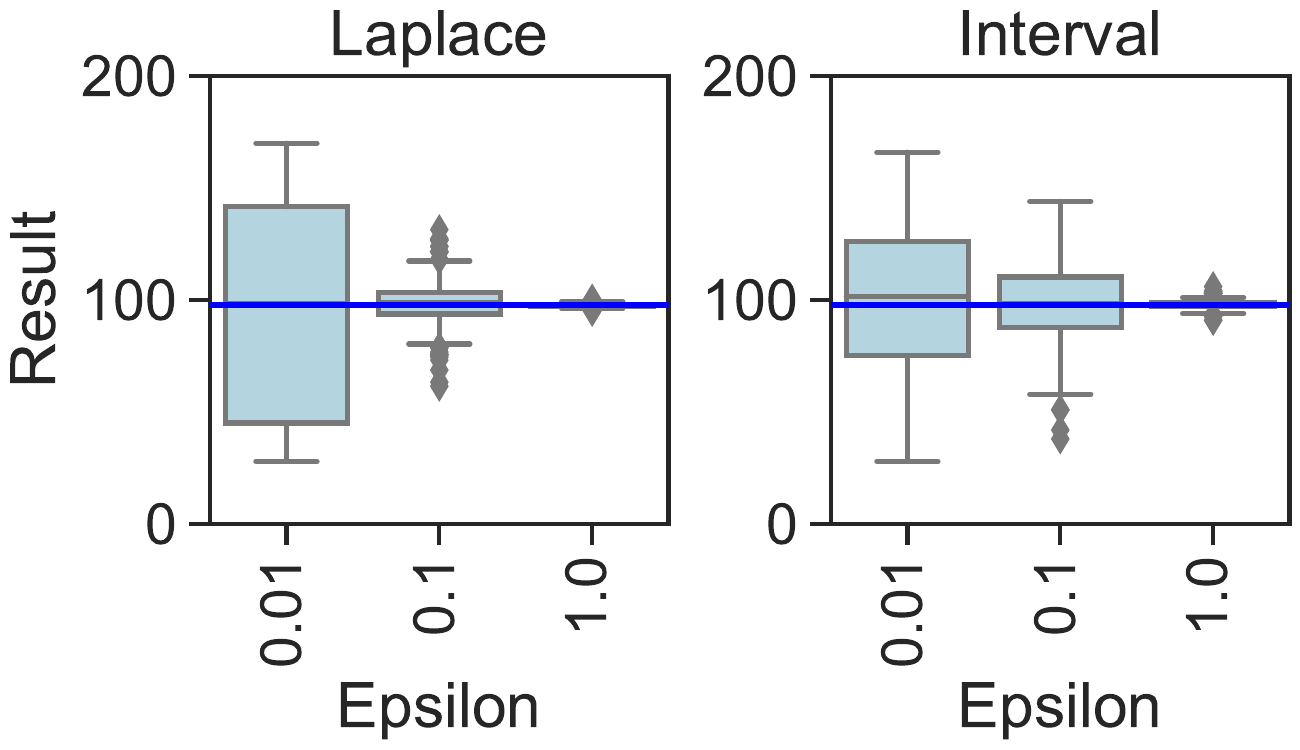}
	\caption{Mean}
	\label{fig:epsilon_mean}
	\end{subfigure}
%	\subfloat[Mean\label{fig:epsilon_mean}]
%	{\centering
%		\includegraphics[scale=0.38]{fig/comparison_sum.pdf}
%	}
%	\hfill
	\begin{subfigure}[b]{0.5\textwidth}
	\centering
	\includegraphics[scale=0.38]{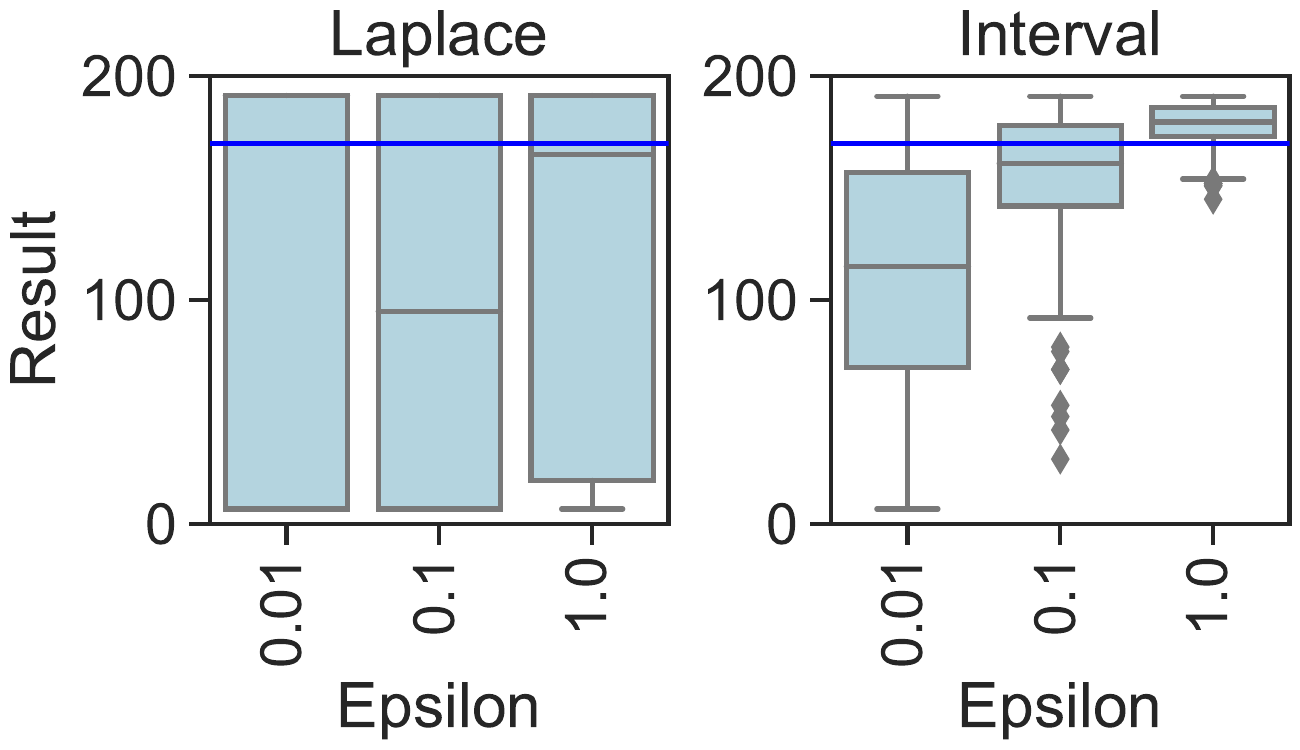}
	\caption{Max}
	\label{fig:epsilon_max}
	\end{subfigure}
%	\subfloat[Max\label{fig:epsilon_max}]
%	{\centering
%		\includegraphics[scale=0.38]{fig/comparison_max.pdf}
%	}
	\vspace{-2em}
	\label{fig:epsilon}
	\caption{Sensitivity of mean and maximum function towards $\epsilon$.}	
	\vspace{-.5em}
\end{figure}

\medskip
\noindent
\textbf{{Epsilon.}}
The results obtained when changing the privacy parameter $\epsilon$ are 
shown in \autoref{fig:epsilon_mean} for $mean$. Both the Laplace and 
interval-based mechanism show a similar increase in the introduced noise. The  
Laplace mechanism yields better results for larger~$\epsilon$.

For $min$ and $max$, however, the interval-based mechanism clearly outperforms 
the Laplace mechanism for all values of $\epsilon$, see 
\autoref{fig:epsilon_max} for the maximum function. Here, the large sensitivity 
for the Laplace mechanism completely obfuscates the actual result $f(X)$, rendering 
the mechanism inappropriate for these functions.

\begin{figure}[!t]
	\centering
	\includegraphics[scale=0.38]{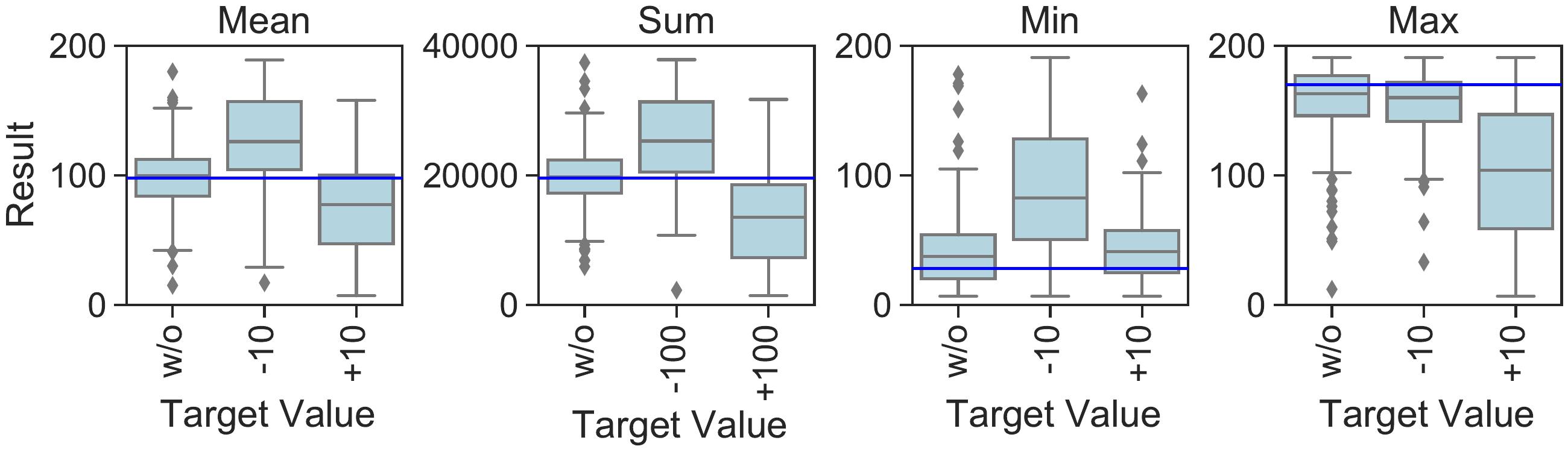}
	\vspace{-.9em}
	\caption{Results for the threshold-sensitive mechanism using differing 
	result thresholds.}
	\label{fig:target_value}
	\vspace{-2em}
\end{figure}

\medskip
\noindent
\textbf{{Threshold-sensitive Mechanism.}}
For the extension of the interval-based mechanism that aims to preserve the 
significance for thresholds, the general trends remain unaffected. However, the 
threshold-sensitive mechanism shifts 
large portions of the probability mass of the output space, as shown in 
\autoref{fig:target_value}. Here, the 
threshold to preserve is $\phi(x): x < f(X)\pm y$, with $y$ being $100$ for 
$sum$ and $10$ for the other aggregation functions. For comparison, the results 
for the interval-based mechanism without threshold preservation are also given. 
There is a clear shift in output probabilities, depending on which values 
preserve the same properties as $f(X)$. Note that the results should not be 
interpreted in absolute terms, but serve as a binary indicator regarding the 
threshold. 
%whether the desired property 
%has been reached or not.

\medskip
\noindent
\textbf{\textbf{Derived Measures.}}
The sample-and-aggregate mechanism for derived measures mirrored the 
trends of the Laplace mechanism for $mean$. This is expected since the 
mechanism is based on the privatized mean. Yet, due to the use of $m$ 
buckets of size $n$, the magnitude of noise is larger. The mechanism 
requires $m$ times as many values in $X$ to achieve the same sensitivity as the 
mechanism for the $mean$. Since the mean is computed using $n$ values per 
bucket, the result estimation is accurate only for large datasets.

\subsection{Case Study: Process for Sepsis Cases}
\label{subSec:caseStudy}

\begin{figure}[!h]
\centering
	\centering
	\includegraphics[width=0.49\columnwidth]{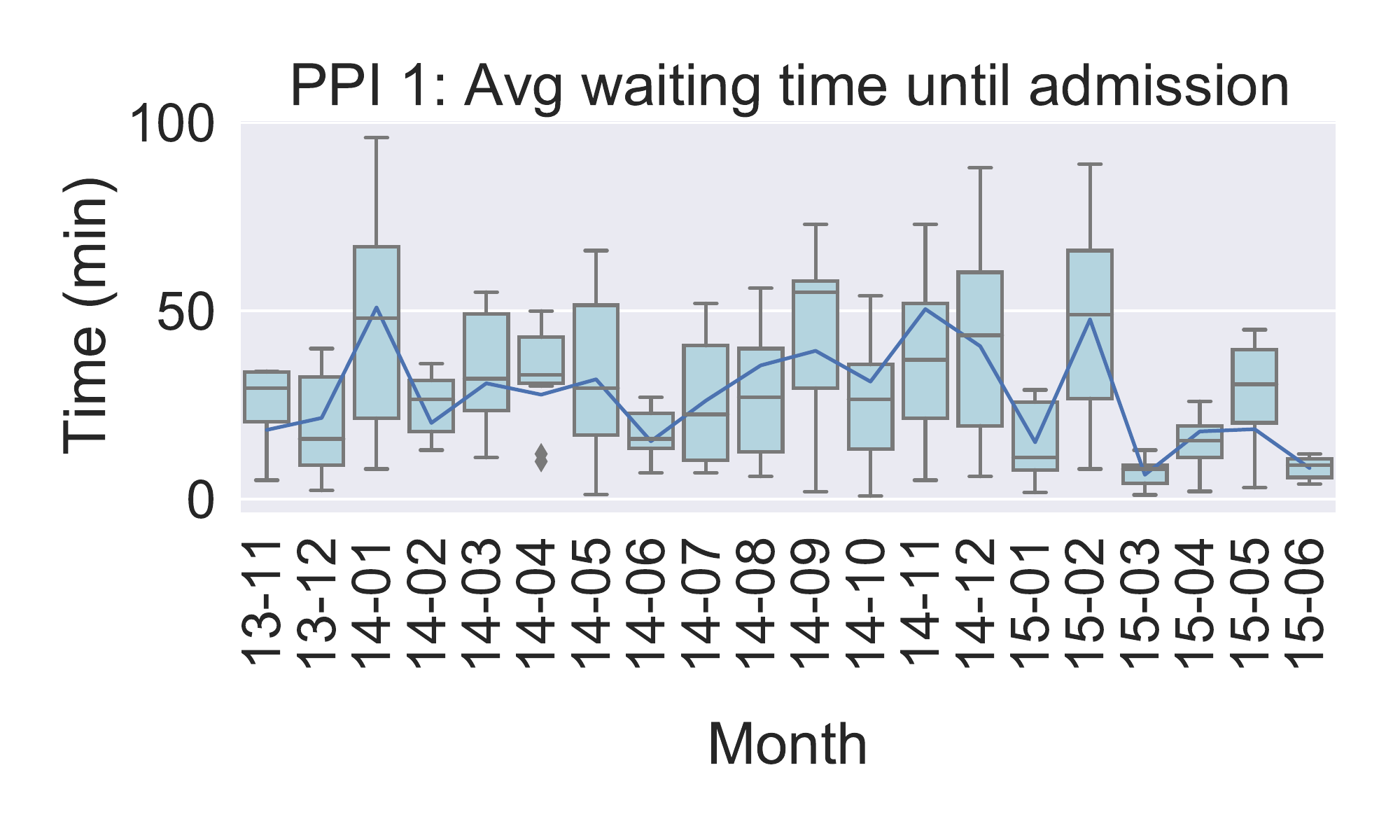}
	\includegraphics[width=0.49\columnwidth]{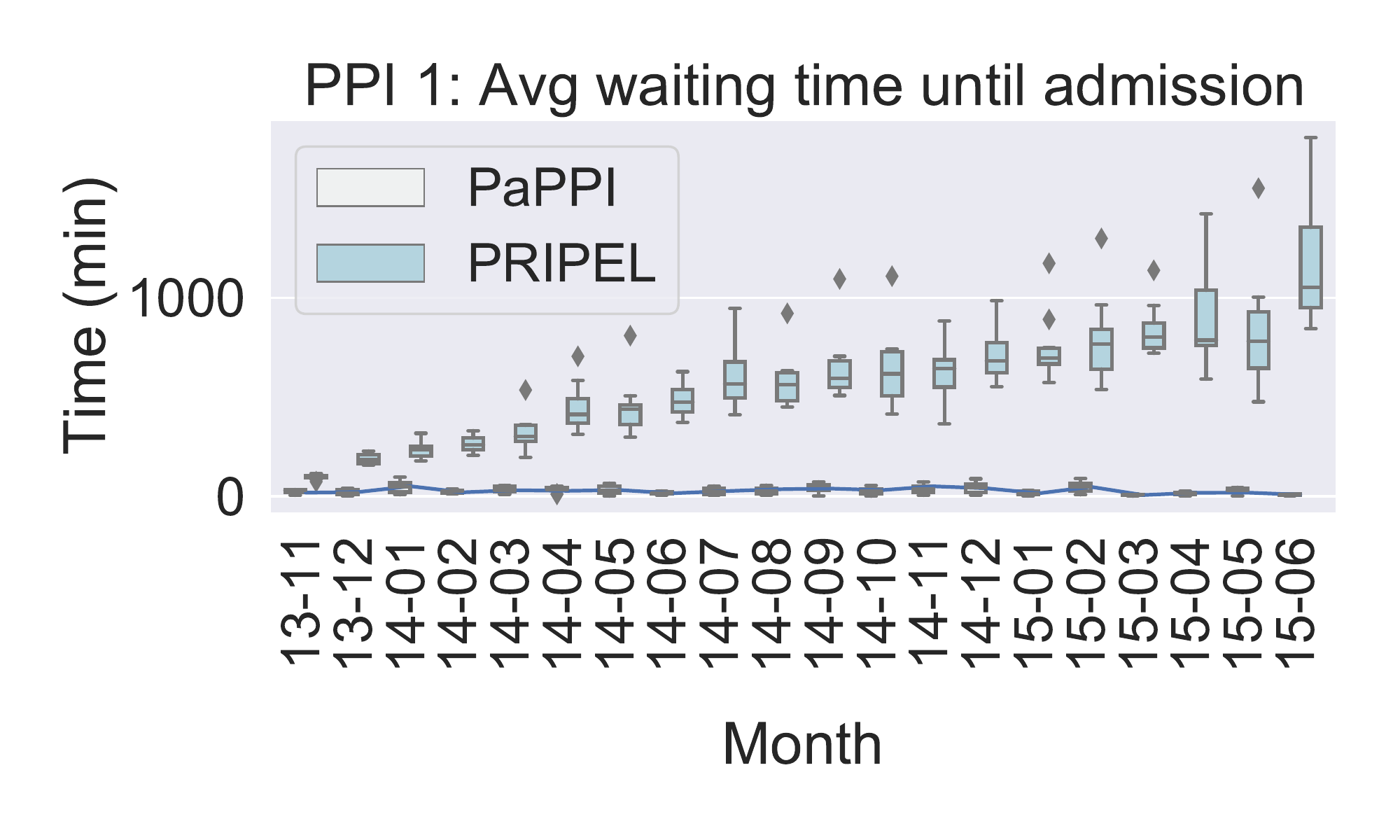}
	\vspace{-1.4em}
	\caption{Evaluation Results for PPI1, only PaPPI (left), and PaPPI and PRIPEL (right).}
	\label{fig:PPI1}
%\subfloat[PPI 6]{
%	\centering
%	\includegraphics[width=0.49\columnwidth]{fig/ppi6_pappi.pdf}
%	\includegraphics[width=0.49\columnwidth]{fig/ppi6.pdf}
%	\label{fig:PPI6}
%}
%	\caption{Analysis results for PPI 1 (a) and PPI 6 (b)}
%	\label{fig:PPIs}
	\vspace{-2.25em}
\end{figure}

\begin{wrapfigure}{r}{0.5\columnwidth}
	\vspace{-2.5em}
	\centering
	\includegraphics[width=0.49\columnwidth]{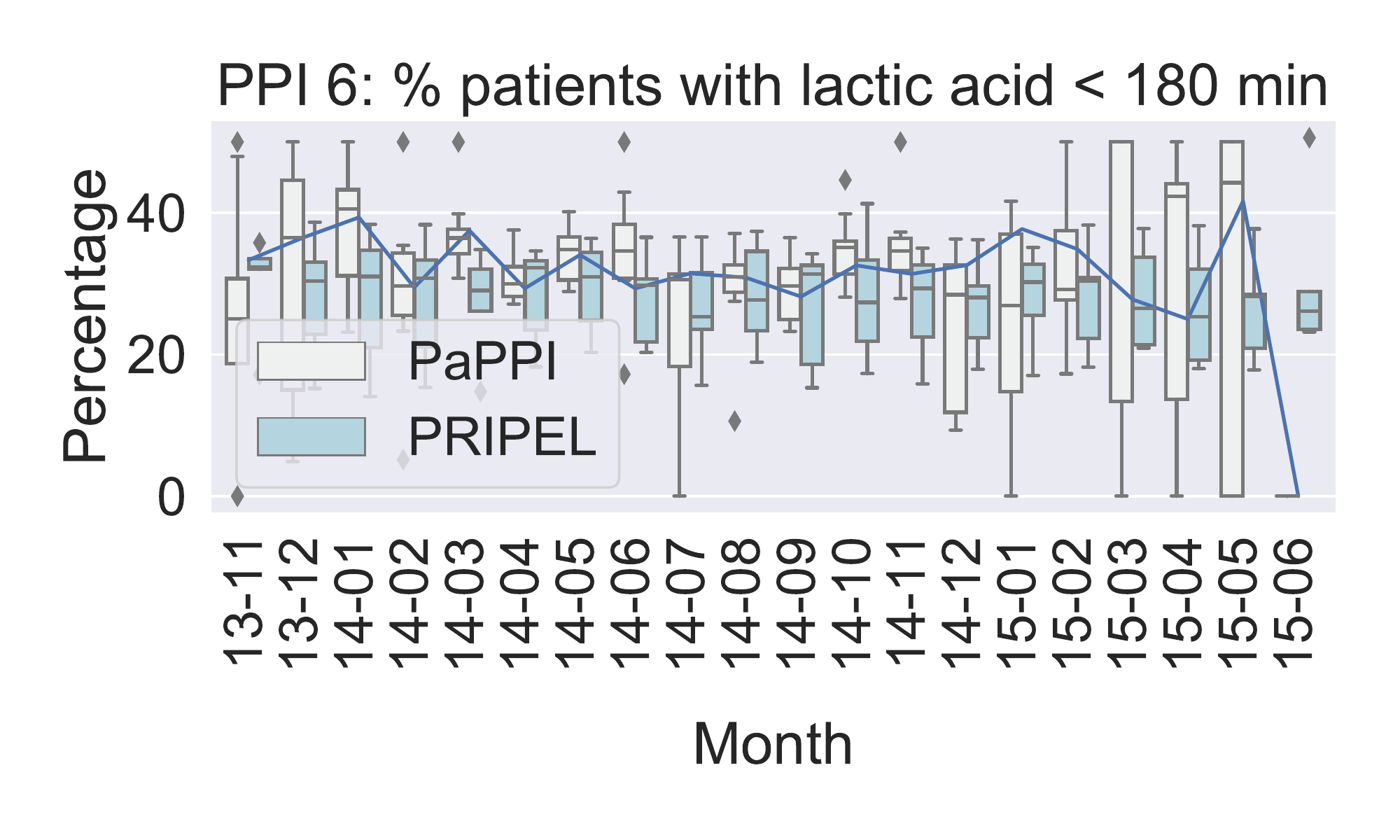}
	\vspace{-1.4em}
	\caption{Evaluation Results of PPI6}
	\label{fig:PPI6}
	\vspace{-2.25em}
\end{wrapfigure}

To explore how the presented mechanisms perform in a real-world application, we 
conducted a study using the Sepsis Cases log. As part of that, we compare 
our approach to a state-of-the-art privatization approach for event logs. That 
is, we evaluated the same PPIs non-anonymously using logs that have been 
anonymized with PRIPEL~\cite{fahrenkrog-petersen2020}.
The PPIs used in our case study were created based on criteria and guidelines presented 
in~\cite{mannhardt2017,stefanini2018} and are listed in 
\autoref{table:PPIs}, together with the employed mechanism used for 
privatization. Some concern the lengths of stays and 
treatments for patients (PPI 1-4), wile others target the adherence to 
treatment guidelines (PPI 5-6). 
%All of these PPIs are evaluated on a monthly scope.
To illustrate the behaviour of our release mechanism, we calculated each 
PPI 10 times using $\epsilon=0.1$ and report aggregate values. 
While results for all PPIs are available online, due to space 
constraints, we here focus on PPI 1 and PPI 6, see \autoref{fig:PPI1} and 
\autoref{fig:PPI6}, respectively.

\begin{table}[!h]
\caption{PPIs defined for the Sepsis Cases log.}
\label{table:PPIs}
	\centering
	\scriptsize
	\begin{tabular}{l  @{\quad} l @{\quad} l l @{\quad}|@{\quad} l}
		\toprule
		ID &  Measure & Target Values & Scope & Mechanism\\
		\midrule
		1:& Avg waiting time until admission & <24 hours& Monthly& Mean - Interval\\
		2:& Avg length of stay & <30 days& Monthly& Mean - Interval\\
		3:& Max length of stay & <35 days& Monthly& Max - Interval\\
		4:& Returning patient within 28 days & <5\%& Monthly& Sum - Laplace\\
		5:& Antibiotics within one hour & >95\%& Monthly& Sum - Laplace\\
		6:& Lactic acid test within three hours & >95\%& Monthly& Sum - Laplace\\
		\bottomrule
	\end{tabular}
	\vspace{-1.5em}
\end{table} 
%\begin{figure}[!t]
%	\centering
%	\includegraphics[width=0.49\columnwidth]{fig/ppi1.pdf}
%	\caption{Analysis results for PPI 1}
%	\label{fig:PPI1}
%\end{figure}
%\begin{figure}[!t]
%	\centering
%	\includegraphics[scale=0.4]{fig/ppi3.pdf}
%	\caption{Analysis results for PPI 4}
%	\label{fig:PPI3}
%\end{figure}

For PPIs 1 to 3, we were able to reconstruct the general trends of the 
non-privatized analysis (exemplified for PPI 1 in \autoref{fig:PPI1}). Yet, we 
also observed specific months with high result variances. For PPI 1 and 2 
($mean$ functions), the 
variance stems from the large domain of input values, resulting in higher 
sensitivities. For PPI 3 ($max$ function), variances were relatively small, 
except for one month, which represents a notable outlier.

The results obtained with our framework are in sharp contrast to those achieved 
when privatizing the event log with PRIPEL before computing the PPIs in a 
regular manner. As shown in \autoref{fig:PPI1} (right), the latter approach 
accumulates an error over the recorded time period. The steadily increasing 
deviation from the true value is caused by traces that represent outlier 
behaviour, which was artificially created by PRIPEL.
% e.g., by putting the admission event at the end of very long traces.

PPIs 4 to 6 were calculated using privatized $sum$ 
functions. Due to the relatively low number of traces recorded per month, the 
application of the sample-and-aggregate-framework for the calculation of the 
final percentage value led to worse results. Here, the 
buckets contained not enough values to approximate the true result well. 
However, using privatized $sum$ functions, the results for PPIs 4 to 6 follow 
the general trends of the true values, see \autoref{fig:PPI6} for PPI 6. 
Similarly, also the computation based on logs privatized with PRIPEL yields 
comparable results. In months, in which few 
traces are selected for a PPI, e.g., at the beginning and end of the 
covered time period, the variance is notably larger for our proposed framework, 
an effect that is avoided by the approach based on event log privatization.
%so that the event log sanitization outperforms the proposed framework.
%containing not more than two values. 
%Note, that the last 4 months do not 
%contain more than 10 values, resulting in unreliable analysis results.

Turning to research question \textit{RQ3}, our results provide evidence that 
the proposed framework enables the computation of privacy-aware PPIs that 
mirror the general trends of their true values. Only for time periods, in which 
the PPI computation is based solely on a few traces, our framework does not 
yield sensible results. Thus, given a sufficiently large number of traces as 
the basis for the evaluation of PPIs, we can expect our framework to retain the 
trends. 
%Additionally, we shown that our approach 
%outperforms the state-of-the-art PRIPEL approach for some PPIs.

	\section{Related Work}\vspace{-.5em}
	\label{sec:relatedwork}
	%Much work on the topic of PPI definitions has been published. 
%PPIs are 
%defined using predicate logic, thus enabling formal verification of the 
%defined 
%PPIs. 
For a general overview of privacy-preserving data mining, as mentioned in 
\autoref{sec:introduction}, we refer to \cite{mendes2017} and 
\cite{aldeen2015}. However, data anonymization commonly 
leads to a trade-off between the strength of a privacy guarantee and a loss in 
data utility. This calls for anonymization schemes that minimize the accuracy 
loss of PPI queries, so that management may still assess the fulfilment of 
operational goals, while the privacy of involved individuals is protected.

To define PPIs, it was suggested to rely on ontology-based 
systems~\cite{wetzstein2008} or resort to predicate logic to enable formal 
verification~\cite{popova2010}. In this work, we followed the PPINOT 
meta-model, which is very expressive due 
to its compositional approach. The compositionality is also the reason why we 
opted for the adoption of differential privacy in our approach. 
%We showed that the function composition trees also lends themselves to the 
%application of differential privacy, as shown in we opted to base our 
%mechanisms on 
%this model.
Other privacy models include k-anonymity~\cite{sweeney2002} and its 
derivatives~\cite{machanavajjhala2006,LiLV07}, which statically mask 
recorded data points. Yet, since the evaluation of PPIs is driven by queries 
and processes continuously record data, these techniques are not suitable. 

In the context of data-driven business process analysis, the re-identification 
risk related to event data was highlighted empirically in~\cite{voigt2020}. To 
mitigate this risk, various directions have been followed, including the 
addition of noise to occurrence frequencies of activities in event 
logs~\cite{mannhardt2019}, transformations of logs to ensure $k$-anonymity or $t$-closeness
 before publishing them~\cite{fahrenkrog-petersen2019,RafieiWA20}, and the 
adoption of secure multi-party computation~\cite{elkoumy2020}.
However, since these 
approaches focus on the control-flow perspective of processes, they cannot be 
employed for the privacy-aware evaluation of PPIs in the general case. 
%Despite recent efforts to 
%extend process-related privatization beyond control-flow, e.g., covering role 
%assignments~\cite{DBLP:conf/simpda/RafieiWA19, DBLP:conf/bpm/RafieiA19}, our 
%framework is the first specifically designed for privacy-aware analysis of process performance 
%characteristics using a rich set of measures.
%The MWEM algorithm proposed in~\cite{hardt2010} overcomes 
%this problem by publishing a differentially private data table. Still, it 
%assumes that the analyst can enumerate the stated queries posed on the dataset 
%prior to privatizing the data set. As these queries and their used data 
%dimensions are unknown beforehand for the PPI analysis use case, this 
%methodology does not achieve the needed flexibility for proper useage contexts.

	\section{Conclusion}\vspace{-.5em}
	\label{sec:conclusion}
	In this paper, we proposed the first approach to privacy-aware evaluation of 
process performance indicators based on event logs recorded during the 
execution of business processes. We presented a generic framework that includes 
an explicit interface to serve as the single point of access for PPI 
evaluation. In addition, for PPIs that are defined following the PPINOT 
meta-model, we showed how to design release mechanisms that ensure 
$\epsilon$-differential privacy. 
%The presented protection model introduces a 
%PPI interface, which serves as the single access points for PPI evaluation 
%requests on recorded data. We derive multiple differentially private release 
%mechanisms tailored to PPI definitions in PPINOT. 
We evaluated our mechanisms on both synthetic data and in a case study using 
the Sepsis Cases log. The results highlight the feasibility of our 
approach, given that a sufficiently large number of process executions is 
available.

In future work, we aim to extend the evaluation of the mechanisms, in order to 
recommend which functions to privatize, for a given 
function tree. This would aid process analysts in receiving PPI results, with 
minimal quality loss. 
%Furthermore, we aim at deriving other mechanisms for the 
%privatization of aggregated and derived measures, that outperform the 
%currently proposed mechanisms in terms of result accuracy.
%This protection model serves as a first step towards the 
%incorporation of privacy-protected PPI evaluations in real usage context.
	
%	\todo{revise bibliography}
	\bibliographystyle{splncs04}
%	\bibliography{refs}

\end{document}